\documentclass[aps,prl,twocolumn,superscriptaddress]{revtex4}

\usepackage{graphicx}

\begin{document}


\title{Highly anisotropic magnon dispersion in Ca$_2$RuO$_4$: evidence for strong spin orbit coupling}


\author{S. Kunkem\"oller}
\affiliation{$I\hspace{-.1em}I$. Physikalisches Institut,
Universit\"at zu K\"oln, Z\"ulpicher Str. 77, D-50937 K\"oln,
Germany}

\author{D. Khomskii}
\affiliation{$I\hspace{-.1em}I$. Physikalisches Institut,
Universit\"at zu K\"oln, Z\"ulpicher Str. 77, D-50937 K\"oln,
Germany}

\author{P. Steffens}
\affiliation{Institut Laue Langevin, 6 Rue Jules Horowitz BP 156, F-38042 Grenoble CEDEX 9, France}

\author{A. Piovano}
\affiliation{Institut Laue Langevin, 6 Rue Jules Horowitz BP 156,
F-38042 Grenoble CEDEX 9, France}

\author{A. Nugroho}
\affiliation{Faculty of Mathematics and Natural Sciences, Institut
Teknologi Bandung, Jl. Ganesha 10, Bandung 40132, Indonesia}

\author{M. Braden}\email[e-mail: ]{braden@ph2.uni-koeln.de}
\affiliation{$I\hspace{-.1em}I$. Physikalisches Institut,
Universit\"at zu K\"oln, Z\"ulpicher Str. 77, D-50937 K\"oln,
Germany}




\begin{abstract}
The magnon dispersion in Ca$_2$RuO$_4$ has been determined by
inelastic neutron scattering on single crytals containing 1\% of
Ti. The dispersion is well described by a conventional Heisenberg
model suggesting a local moment model with nearest neighbor
interaction of $J$=8\,meV. Nearest and next-nearest neighbor
interaction as well as inter-layer coupling parameters are
required to properly describe the entire dispersion. Spin-orbit
coupling induces a very large anisotropy gap in the magnetic
excitations in apparent contrast with a simple planar magnetic
model. Orbital ordering breaking tetragonal symmetry, and strong
spin-orbit coupling can thus be identified as important factors in
this system.

\end{abstract}

\pacs{7*******}


\maketitle

The properties of strongly correlated systems with significant
spin-orbit coupling (SOC) present a challenging problem.
The intensively studied example is the reduction of the magnetic
state of Ir$^{4+}$ (electronic structure $5d^5$ or $t_{2g}^5$,
$L_{eff}$ = 1, S=1/2) to an effective Kramers doublet with $J$=1/2
\cite{1}. But even more drastic effects can be expected for heavy
ions with $d^4$ occupation ($t_{2g}^4$, $L_{eff}$=1, S=1), e.g. in
Ir$^{5+}$, Ru$^{4+}$, Os$^{4+}$ etc.\cite{2}. According to Hund's
rules (generalized for ions sensing crystal electric fields) the
ground state should be a nonmagnetic singlet with $J$=0, see e.g.
reference \cite{3,4}. And indeed isolated Ir$^{5+}$ ions and also
most of concentrated Ir$^{5+}$ compounds are nonmagnetic, although
a few magnetic Ir$^{5+}$ cases are known \cite{5}. In a solid
magnetic order can occur even if the ground state of an isolated
ion is a singlet, see chapter 5.5 in reference  \cite{4}, but it
requires a strong exchange interaction, so that the exchange
splitting of excited magnetic states (in the Ru$^{4+}$ case a
$J$=1 triplet) is larger than the energy difference between the
ground-state singlet and the excited triplet, which is given by
the SOC parameter $\lambda$. The SOC can also be at least
partially suppressed by a non-cubic crystal field (CF),
$\Delta_{noncub}$, which splits the $t_{2g}$ ($L_{eff}$=1) triplet
and stabilizes real orbitals. Both these factors, CF and magnetic
interaction, can combine to suppress the $J$=0 state and to
eventually induce the magnetically ordered ground state. In terms
of energy scales, one should expect such magnetic ordering for
$\Delta_{noncub} + \mu \cdot H_{exch}
> \lambda$, which seems quite unlikely for Ir$^{5+}$  where
$\lambda$=$\frac{\zeta}{2S} = \zeta/2$ amounts to 0.2 to 0.25 eV
($\zeta$ is the atomic spin-orbit parameter). But for $4d$
compounds this relation can easily be reached, as for Ru$^{4+}$
$\lambda \sim 0.075$\,eV\cite{2,7}.
Indeed, practically all Ru$^{4+}$ compounds order magnetically
aside from the metallic ones - and even some metallic ruthenates
are magnetic, such as the ferromagnetic metal SrRuO$_3$. The
persisting role of SOC in these magnetic Ru$^{4+}$ compounds is an
intriguing open issue.

\begin{figure}
\includegraphics[width=0.99\columnwidth]{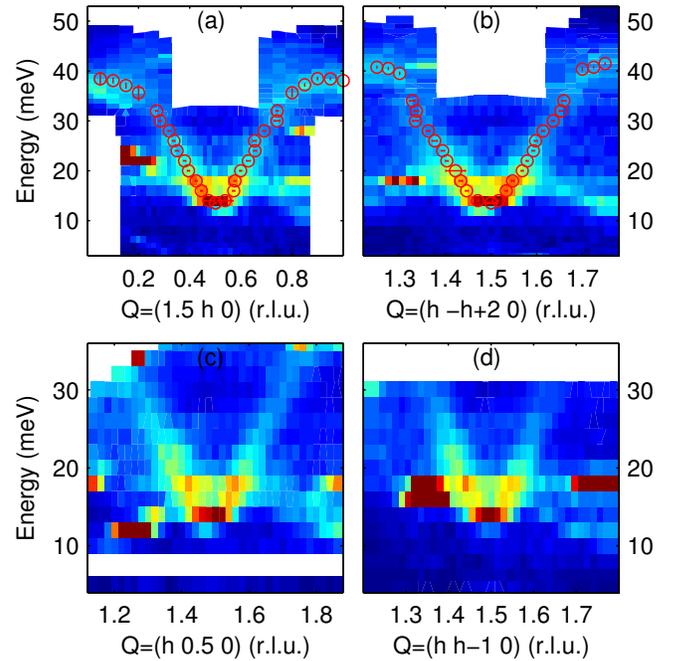}
\caption{\label{hk0} Intensity distribution in energy versus
scattering vector, $\mathbf Q$, planes taken at 2\,K around the
(1.5\,0.5\,0) magnetic zone center. a) and c) show the
symmetrically
 equivalent dispersion along (0,$\xi$,0) and ($\xi$,0,0) direction, subplot b) and d)
along ($\xi$,$\xi$,0) and ($\xi$,-$\xi$,0). The color coding
corresponds to the raw data. Open symbols indicate the dispersion
obtained by fitting single scans. Data were taken with final
energies of 35\,meV for constant $\mathbf Q$ scans at high energy
transfer and 14.7\,meV elsewhere.}
\end{figure}

Ca$_2$RuO$_4$ (CRO) is such a Ru$^{4+}$ case, which has been
intensively studied as the Mott-insulating analogue of the
unconventional superconductor Sr$_2$RuO$_4$\cite{8,8b,9,10}. CRO
exhibits a metal-insulator (MI) transition at 357\,K which is
accompanied by a flattening of the RuO$_6$
octahedra\cite{9,10,11,Pbca}. This flattening continues upon
further cooling until it saturates near the onset of magnetic
order at T$_N$$\sim$110\,K. The magnetic structure is
antiferromagnetic (AFM) with moments aligned parallel to the
layers \cite{9,Pbca}. The electronic structure has been studied by
various approaches \cite{12,13,14,15,25}. From the spectroscopic
study of CRO it was concluded that SOC indeed plays an important
role but is not sufficiently strong to stabilize the $J$=0 state
\cite{12}. Density functional theory calculations indicate a
pronounced shift in orbital polarization leading to almost full
occupation of the $d_{xy}$ levels at low
temperature\cite{13,14,15,25,15b}. More recently the $J$=0 state
was explicitly proposed for CRO \cite{2,7}. Starting from the
scenario of strong SOC and including noncubic CF and intersite
exchange, the magnetically ordered state in CRO is reproduced and
several unusual features of the magnetic excitation spectrum of
CRO are predicted, such as a peculiar shape and large width.
The alternative, more conventional picture is to attribute the
magnetism of CRO to  the conventional S$\sim$1 state of Ru$^{4+}$
ions, with SOC playing a less significant but still prominent
role. In this case one can describe the magnetic state, including
spin waves, by the usual exchange Hamiltonian.

Here we present an inelastic neutron scattering (INS) study and
spin-wave calculations of the magnetic excitations in CRO. We find
that a conventional model can well describe the obtained
dispersion, while there are considerable differences with the
proposed $J$=0 model\cite{7}. Most interestingly there is a
sizable spin-gap which indicates that rotating the magnetic moment
within the layers costs large energy. The breaking of the local
tetragonal symmetry and the associated orbital polarization, which
has been neglected in theory so far \cite{13,14,15,25,15b}, are
important parameters to understand the magnetism in CRO.

\begin{figure}
\includegraphics[width=0.99\columnwidth]{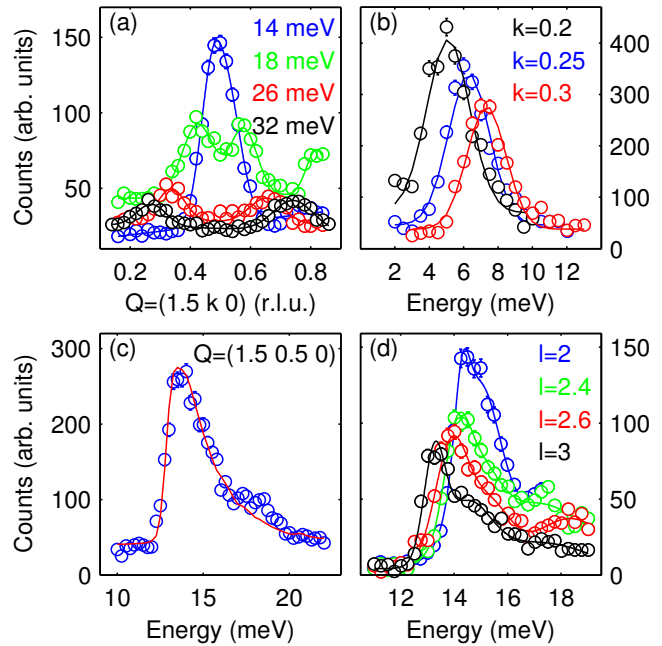}
\caption{\label{scans} Several characteristic scans: a) Constant
energy scans at (1.5,$k$,0) taken at 2\,K fitted with gaussians
and background. b) Phonon scans taken at $\mathbf Q$=($\xi$,2,0);
the lines correspond to the folding of the resolution function
with a simple linear phonon dispersion. No additional parameter
is needed to describe the shape of the intensity profile. c) and
d) show the energy scans at the zone center (1.5 0.5 0) and at
(0.5 0.5 $l$).}
\end{figure}

CRO single crystals of of several 100mm$^3$ volume containing 1\%
of Ti were obtained by the travelling solvent floating zone method
in a mirror furnace (Canon SC1-MDH11020-CE).
We added 1\% of Ti as this seems to avoid bursting of the crystal
upon cooling below the  MI transition.
Characterization by magnetic susceptibility and by neutron
diffraction experiments indicate a magnetic transition at
T$_N$=112\ K and no significant impact of the very small amount of
Ti. INS experiments were performed with the IN8 triple-axis
spectrometer.
Most experiments were performed with fixed final
momentum of $k_f$=2.662\,\AA$^{-1}$ (E$_f$=14.7\,meV); some scans
at high energy transfer or aiming at better resolution were
performed with $k_f$=4.1 and $k_f$=1.97\,\AA$^{-1}$, respectively.
We studied the magnon dispersion in the two scattering geometries
(100)/(010) and (110)/(001) in reduced units of the tetragonal
lattice \cite{Pbca}. For both set\-ups two crystals were
coaligned. The sample of the second scattering plane was
essentially untwinned as determined on the IN3 spectrometer.

Fig. 1 shows color mappings of the measured intensity
distribution. Due to the weakness of scattering in CRO (small
moment and rapidly decreasing form factor) contaminations by
various phonon branches are highly visible. By analyzing and
comparing results taken in different Brillouin zones and
geometries the dispersion can be unambiguously determined. Magnon
excitations start at the AFM Bragg points ($\frac{2n_h+1}{2}\
\frac{2n_k+1}{2}\ n_l$) with integer $n_h$, $n_k$ and $n_l$.
However, there is a sizeable spin gap of 13.04(5)\,meV. For a
square planar  antiferromagnet the magnon dispersion extends from
$\mathbf Q$=(0.5 0.5) to (0.75 0.75) in [1 1] direction, as (1 1)
is a Bragg point, and to (0 0.5) in [1 0] direction. $\mathbf
Q$=(0.25 0.25) and (0 0.5) are AFM Brillouin zone boundaries. In
CRO there is, however, a severe structural distortion \cite{Pbca}.
Some characteristic scans performed to determine the magnon
dispersion in CRO are shown in Fig. 2. Constant energy scans at
intermediate energy cut through the magnon cones at two positions.
Constant $\mathbf Q$ scans taken just at the AFM zone center show
a characteristic asymmetric shape, see Fig. 2 (c) and (d):
Intensity rapidly increases when crossing the spin gap and slowly
diminishes with further energy increase. We have calculated the
folding of the spin-wave dispersion including its expected signal
strength with the experimental resolution using the RESLIB
\cite{16} package and verified that scans across transversal
acoustic phonons are well reproduced, see Fig. 2 (b). The steep
spin wave dispersion perfectly describes the asymmetric shape of
the spectra taken at the zone center, see Fig. 2 (c). The total
width of the dispersion is low, as maximum energies of 37.8(3) and
41.2(5)\,meV are reached at the magnetic zone boundaries,
(0.5\,0\,0) and (0.25\,0.25\,0).

In order to describe the magnon dispersion we use a conventional
Heisenberg model with a single-site anisotropy term
arising from SOC:
$H =
\sum_{\mathbf{i,j}}\mathbf{J}_{i,j}\mathbf{S_i}\cdot\mathbf{S_{j}}-\delta
\sum_i  (S^y_{i})^2$.
We set S=0.67 following the neutron diffraction study\cite{9}. The
sum runs over pairs of magnetic ions, so that each pair or bond
appears twice. Spin waves were calculated with the
Holstein-Primakoff transformation as described in references
\cite{17,18}. We include the nearest-neighbor magnetic exchange of
J=8\,meV, next-nearest neighbor interaction along the orthorhombic
$a$ and $b$ directions of J$_{nna}$=J$_{nnb}$=0.7\,meV and an AFM
coupling between neighboring layers. The next-nearest neighbor
interaction is chosen isotropic, as the twinned crystal used in
the (1\,0\,0)/(0\,1\,0) geometry prohibits distinguishing these
directions. The need for the additional parameter can be seen when
comparing the magnon energies  at $\mathbf q$=(0.25 0.25 0) and
(0.5 0 0), which are identical in the model with only
nearest-neighbor interaction. The interlayer coupling,
J$_c$=0.03\,meV is the only parameter that breaks the tetragonal
symmetry in our model aside from the single-ion anisotropy. Note,
however, that the crystal structure is orthorhombic lifting the
degeneracy of magnetic interaction parameters. We chose the AFM
interaction between the Ru at (0,0,0) and that at (0,0.5,0.5) (in
the orthorhombic cell \cite{9}), which stabilizes an $A$ centered
magnetic structure with magnetic space group Pbca \cite{Pbca}.

\begin{figure}
\includegraphics[width=\columnwidth]{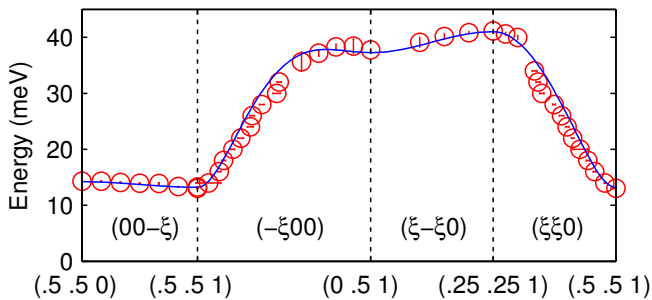}
\caption{\label{dispersion} (Dispersion of the magnon branch along
the main symmetry directions at T=2\,K. The open symbols indicate
the values obtained by fitting the raw data scans with Gaussians
or by folding the resolution function with the modeled
dispersion. Lines correspond to the spin-wave calculations with
the Heisenberg model as described in the text.}
\end{figure}

The magnetic moment in CRO points along the orthorhombic $b$
direction. Therefore, one might  expect a large gap for the
magnetic excitations involving rotations of the moment out of the
RuO$_2$ layers, and much softer in-plane modes. The latter are
described by the expectedly small in-plane anisotropy. Following
reference \cite{18,qureshi} both branches can be described
simultaneously with two anisotropy parameters. Surprisingly, in
CRO the in-plane anisotropy turned out to be extremely strong. The
magnon dispersion starts at 13.04(5)\,meV which we may identify
with the in-plane gap. There is no magnon branch at lower energy
as is clearly shown in the intensity maps. We find some weak
intensity at an energy transfer of 5\,meV appearing near $\mathbf
Q$=(1.5\,0.5\,0), but this signal is flat in energy and restricted
to the AFM zone-center. Furthermore, this signal is much weaker
than the magnon modes at higher energies suggesting a possible
origin associated with the Ti impurities and some local effect. As
shown in Fig. 2 (d) and 3 there is a finite inter-layer dispersion
visible in the scans taken at $\mathbf Q$=(0.5\,0.5\,$q_l$) with
the second untwinned crystal. The tetragonal [110] direction
corresponds to orthorhombic $b$ in the used mounting and thus to
the direction of the magnetic moment; therefore, the transverse
magnon with in-plane polarization (thus parallel to orthorhombic
$a$) fully contributes. Also in the other configuration there is a
clear difference in spectra taken at $\mathbf Q$=(0.5\,0.5\,0) and
=(1.5\,0.5\,0). For the twinned sample we superpose AFM
zone-centers and zone boundaries. For a twinned crystal $c$
polarized magnons will always contribute, while for the in-plane
magnon the geometry condition that only magnetic components
perpendicular to $\mathbf Q$ contribute, suppresses some modes.
The fact that we see a  clear difference at various
($\frac{2n_h+1}{2}\ \frac{2n_k+1}{2}\ n_l$) unambiguously shows
that the modes dispersing between 13.04(5) and 14.2(1)\,meV posses
an in-plane polarization. This furthermore agrees with the $Q_l$
dependence of the signal. We may thus conclude that the lowest
magnon branch in CRO possesses an in-plane character and that it
disperses between 13.04(5) and 14.2(1)\,meV along the $c$
direction and up to 41.2 and 37.8\,meV along ($\xi$ $\xi$ 0) and
($\xi$ 0 0) paths, respectively. We cannot identify  the $c$
polarized modes as they may remain hidden in the asymmetric shape.
There is some evidence for a nearly flat branch around 36\,meV,
but we cannot fully rule out that these modes are purely nuclear
or that they carry longitudinal polarization. For simplicity, the
experimental dispersion is described by an easy-axis anisotropy
\cite{qureshi}, see Fig. 3.

The magnon dispersion including its large gap can be very well
described within the spin-wave theory suggesting a conventional
local moment S$\sim$1 magnetism with a strong - but not decisive -
impact of SOC. Starting from the other scenario, a spin-orbit
driven $J$=0 singlet nature which is rendered magnetic by noncubic
CF and intersite exchange,  Akbari and Khaliullin \cite{7}
predicted several unusual features of the magnetic excitation
spectrum, such as the energy continuously softening from the value
$\lambda$ at $\Gamma$, and the presence of extra modes in some
part of the spectrum. Our results, however, do not support this
model \cite{7}. First, the observed dispersion is much flatter
than this prediction, as it does not reach energies of the order
of the expectedly large values of $\lambda$, and as there is a
strong gap. Second, the singlet picture predicts a continuously
increasing dispersion near the AFM zone boundaries, while our
experiments find the saturation predicted by the Heisenberg model,
see Fig. 1 and 2. The Heisenberg scenario also implies several
branches: two transversal branches arise from the orthorhombic
anisotropy (in-plane and $c$ polarized), and longitudinal modes
can exist in CRO being on the border to itinerancy.

Using the standard description, with the hopping parameters
$t\sim$100\,meV, obtained by ab-initio calculations \cite{15c,
19}, and using the Hubbard's $U$$\sim$2\,eV, we would obtain for
the exchange constant J=2$t^2/U\sim$10\,meV, in good agreement
with our experimental finding. However, CRO is not a strong Mott
insulator with completely localized electrons as it is already
indicated by the low-lying MI transition.
In this case the basic $J$=0 Ansatz may be not a good starting
point, as the $J$=0 state can be suppressed by electron hopping.
Also for Ir$^{4+}$ (specifically for Na$_2$IrO$_3$) the sizable
hopping modifies the whole picture \cite{20,21}, leading to novel
quasi-molecular orbital states with reduced impact of SOC. The
conspicuous but typical absence of $J$=0 physics in most of the
Ru$^{4+}$ materials seems largely connected with the hopping.

\begin{figure}
\includegraphics[width=\columnwidth]{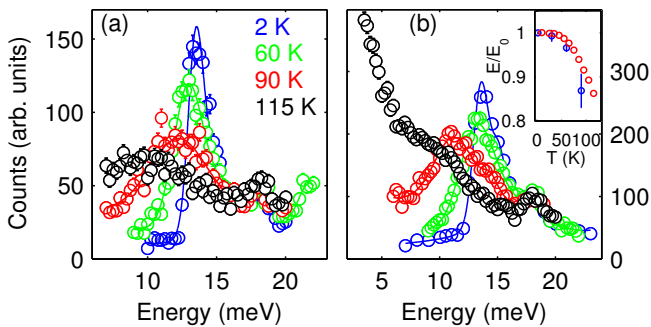}
\caption{\label{tdep} Temperature dependence of the magnetic
scattering at the AFM zone center (1.5,0.5,0) measured with the Si
monochromator and $k_F$=1.97 (a) and 2.662\AA$^{-1}$ (b). The
sharp peak associated with the in-plane spin gap mode softens and
broadens considerably. The inset shows the energy of the spin gap
scaled to its low temperature value (blue) compared to that of a
Raman signal taken from \cite{24} (red).}
\end{figure}

Another argument in favor of the applicability of the usual
picture of Ru$^{4+}$ ions (S$\sim$1) is the strong flattening of
RuO$_6$ octahedra \cite{9,11} occurring below the MI transition.
Such distortion is typical for the usual Jahn-Teller effect: it
stabilizes the doubly-occupied $d_{xy}$ orbital, leaving two
electrons on $d_{xz}$ and $d_{yz}$. In such state the orbital
moment and spin-orbit interaction are partially quenched. The sign
of this distortion proves that in this system the Jahn-Teller
effect is stronger than the SOC which would have caused the
opposite distortion and CF splitting \cite{4}. Recent spectroscopy
data \cite{25} confirm this significant splitting of $t_{2g}$
orbitals.

On the other hand the observation of the strong in-plane magnetic
gap is remarkable for a layered system. It underlines  the
relevance of the SOC in CRO even in the conventional scenario.
Several Raman scattering experiments observed an additional signal
in $B_{1g}$ symmetry appearing in the AFM phase \cite{22,23,24}.
This feature was interpreted as a two-magnon excitation, but our
results clearly rule out such explanation. The Raman feature
appears at 102\,cm$^{-1}$=12.6\,meV at 10\,K which is much below
the energies for two magnon excitations and the expected peak in
the two-magnon density of states (near 80\,meV). Instead this
energy agrees with that of the in-plane gap mode. The single
magnon mode, however, is not Raman active in first approximation,
but SOC can induce a finite signal. The temperature dependence and
the extreme broadening of the Raman signal at higher temperature
agree reasonably well with the corresponding behavior of the
magnon gap, see inset in Fig. 4 (b).

The magnetic in-plane anisotropy in CRO must originate from SOC
and from an orbital arrangement breaking tetragonal symmetry.
There have been many experimental and theoretical analyzes
\cite{12,13,14,15,25,15b} elucidating the change of the orbital
polarization upon cooling and the increasing occupation of the
$d_{xy}$ versus the $d_{xz}/d_{yz}$ orbitals following the
flattening of the RuO$_6$ octahedron. This distortion possesses
$E_g$ symmetry, which is the most frequently analyzed in
Jahn-Teller models \cite{bersuker}. The $t_{2g}$ orbitals,
however, also couple to the $T_{2g}$ octahedron distortions
\cite{bersuker} which break tetragonal symmetry in the case of CRO
but which were neglected so far. The temperature dependence of the
crystal structure of CRO in the insulating phase reveals an
ongoing elongation of the RuO$_6$ octahedra \cite{9,11} along the
orthorhombic $b$ direction along which moments align. This
distortion corresponds to the $T_{2g}$ "scissor" mode of the free
octahedron \cite{bersuker} lifting the $d_{xz}/d_{yz}$ degeneracy.
Similar to a tetragonal distortion, e.g. along z-axis, which would
stabilize orbitals with $l^z$ = $\pm$1, $(d_{xz}\pm id_{yz})$, and
which, by SOC would orient spins along z-direction, (or a trigonal
elongation along [111] (in cubic setting), which would make [111]
an easy axis, see e.g. \cite{4}), such $T_{2g}$ distortion
(elongation along $b$ axis) makes the orthorhombic b-direction the
easy axis.

In conclusion we have studied the magnon dispersion in CRO which
considerably differs from recent predictions for a $J$=0 singlet
ground state. Instead, the dispersion is well described in a local
moment Heisenberg model with strong anisotropy terms yielding a
nearest-neighbor exchange interaction of J=8\,meV which agrees
with the large calculated hopping integrals. Large hopping seems
to be the main cause for the suppression of the $J$=0 state in
Ru$^{4+}$ compounds. On the other hand, the remarkably strong
in-plane anisotropy clearly shows that considering tetragonal
crystal fields is insufficient. There is important orbital
polarization breaking tetragonal symmetry, which is related to the
prominent elongation of RuO$_6$ octahedra along the orthorhombic
$b$ direction and which renders spin-orbit coupling still active
in this system.

\begin{acknowledgments}
Part of this work was supported by the Institutional Strategy of
the University of Cologne within the German Excellence Initiative
and by Deutsche Forschungsgemeinschaft through project FOR 1346.
We acknowledge stimulating discussions with M. Gr\"uninger, G.
Khaliullin, P. van Loosdrecht, and T. Lorenz.

\end{acknowledgments}


\begin{thebibliography}{9}
\bibitem{Pbca}
CRO crystallizes in space group Pbca with lattice constants
$a$=5.39, $b$=5.63 and $c\approx12$\AA. We use the tetragonal
lattice in space group I4/mmm with $a=b\approx3.9$\AA \ and
$c\approx12$\AA \ in accordance with most publications. Spin-wave
calculations were performed in the orthorhombic cell and
transformed to the tetragonal one.
\end{thebibliography}
\end{document}